**Beyond Attoseconds**

*A. E. Kaplan*

Department of Electrical and Computer Engineering
The Johns Hopkins University, Baltimore, MD 21218

**Abstract**

We briefly review the pilot ideas on the generation of EM-pulses much shorter than already available sub-femtosecond pulses, and outline inroads and venues into the physics of pulses mush shorter than an attosecond ($10^{-18}\,s$), in particular the so called *zeptosecond* ($10^{-21}\,s$) and *yoctosecond* ($10^{-24}\,s$) pulses that may allow one to operate on QED and nuclear as well as quark-gluon time plasma scales. We also very briefly outline the entire time-scale available in the existing universe, down to the ultimately short the so called ***Planck*** time $\sim 10^{-43}\,s$, which is the time-scale of Big Bang, and the most significant time-scale-posts on the road to it.

Generation of very short coherent pulses with high repetition rates has a great importance for applications both in fundamental and applied physics. Our quest for ever shorter pulses is relentless if not as fast as we wished it to be. The end of 80-ties, about 30 years after the invention of a laser, saw optical pulse as short as 6 to 5 *fs* based on the pulse compression technique [1]. The next in line were sub-femtosecond and attosecond pulses, which were reaching into new, sub-cycle domain with huge spectral width. The main technique of choice to go there was to use very high harmonics generation (HHG) by a regular short laser pulse, which was first discovered by the group of C. K. Rhodes [2] in 1981, who observed up to 46-th order harmonics of a $10.6 - \mu m\ CO_2$ laser with intensities greater than $10^{15}\ W/cm^2$ in inertially confined plasma. Based on this new phenomenon, the feasibility of generating sub-femtosecond and attosecond pulses by using HHG in noble gases have been discussed in [3-6] by various groups in the beginning of 1990-ties. Amazingly, a title of [4] laid down all the major features of the idea and the notion of attosecond pulses: "...attosecond light-pulse generation using laser-induced multiple-harmonic conversion processes in rare-gasses". The super-short pulses are generated by the comb of higher harmonics the same way as short pulses in a laser with many phase-locked modes, but on a vastly broader scale of spectrum. The current state-of-art in the field is impressive: first ground-breaking results were reported generation of *as*-pulses in the sub-femtosecond domain (longer than $100\ as$ [7,8]); the latest report [9] suggests experimental evidence of *as*-pulses down to ~ *20 as*.

Other ideas of generating those pulses were based on different nonlinear-optics effects. It is worth to mention here another avenue [10] that relied on the generation of the comb of equidistant frequencies using the Cascade Stimulated Raman Scattering (CSRS), whereby laser light with the frequency $\omega_L$, propagating in the Raman-active medium with the Raman frequency $\omega_0 \ll \omega_L$, excites many cascade-induced Stokes and anti-Stokes components with the frequencies $\omega_j = \omega_L + j\omega_0,\ j = \pm 1,\ \pm 2,\ \pm 3...$ . CSRS (known for many years [11] provides tremendously broad spectrum, with up to ~10-15 CSRS lines being spread from far infrared (down do a few microns) to very far ultraviolet and carrying a considerable power due to very high conversion efficiency. If all these CSRS components were properly phased or locked to each other, this broad spectrum radiation could transform itself into the train of SFP pulses, with each of individual pulses having the length (time duration) of the order of one cycle of the highest-frequency component (which can be much shorter the the cycle length of the pump radiation), and extremely high intensity. In the work [10] on generating pulses (and related *2π*-solitons) down to ~ *0. 2 fs = 200 as* it was argued that the technique may have great improvement over HHG approach, since the CSRS comb may carry up to 40% of pump energy conversed into CSRS lines, as compared to HHG radiation that has a very low efficiency of transformation of pumping laser into HHG comb (typically much less than $\sim 10^{-4}$). While the proposal [10] relied on self-organized CSRS generation, the work [12] proposed to greatly enhance the CSRS generation by using two pumping laser frequencies that would allow for control of the fundamental Raman frequency by putting it into the spectral domain optimal for matching the group velocities of most of the cascade Raman components. The work [13] experimentally demonstrated that generation of single-cycle pulse down to ~ *1 fs*. One of the directions of further development in the field is to bring together CSRS and HHG approaches [14] to try to reach new limits in the *as* technologies.

Now, what is beyond attoseconds? Of course, when trying to look beyond that horizon, one should not be thinking in terms of sheer numbers and zeros; the main question is -- what is the physics in this race? What new physical phenomena and domains we are trying to reach? In this case it would be instructive to translate the time into highest energy of photons carried by the pulse the time into highest energy of photons carried by the pulse. The highest frequency of the Fourier spectrum, $\omega_{mx}$, of a ***non − oscillating*** pulse is inversely proportional to its duration, $\tau$ as $\omega_{mx} \sim 1/2\tau$. Since that energy is proportional to the frequency with the proportionality coefficient being a Planck constant $\hbar$ as $E_{mx} = \hbar\omega$, so that similarly to the uncertainty principle, whereby $\delta E \times \delta t \geq \hbar/2$, we connect the higher energy of EM-quanta, $E_{mx}$, carried by the pulse, to the time $\tau$ as

$$E_{mx} \sim \hbar/\tau \tag{1}$$

(here we took into consideration that in a ***propagating***, i. e. at least single-cycle pulse, we have $E_{mx} \approx 2 \times \delta E$) The sub-picosecond and femtosecond domain, with the photon energies less than ~ *0. 1 eV*, became a fertile field for research, discoveries and quests for applications, ranging from the registration of super-fast processes, to time-resolved spectroscopy, to the characterization of semiconductors with sub-ps relaxation times; another application is the so-called Terahertz (THz) technology, which uses electromagnetic pulses as a diagnostic tool to "see through" opaque materials and structures. One of the most fundamental breakthroughs with rich potential for practical applications [15] was the discovery and development of chemical reaction control and femtosecond time resolution by using powerful femtosecond laser pulses with molecular-domain photon energies.

Before we go further, a comment regarding the generation of short pulses is in order here. It should be made very clear that when we talk about pulse generation, we mean -- ***controllable, coherent, and reproducible*** pulses. Only those pulses whose timing, temporal profile of the field, its duration and phase are reproducible, can be regarded as controllable generated objects. Regular laser pulses are essentially laser oscillations modulated by relatively slow "envelope".

Ideally, the shortest pulses would have non-oscillating nature during the entire pulse cycle; they have to be  so short that some of them are just a single burst of rising-and-falling electrical field (which is a so called "half-cycle" pulse that, upon propagation, becomes a "single-cycle" pulse due to diffraction of the related super-broad spectrum; the result of that is the formation of "time-derivative" of the original pulse, when it is observed into far-field area [16]).  The Fourier spectrum of such a pulse is reminiscent to that of a black-body radiation, but with a huge difference: in the case of a single half-cycle all the spectral components at different frequencies have ideally the same phase, which can be described as a "trans-spectral coherence" across the entire super-broad spectrum, a feature never encountered in a regular laser optics.  Any modulation or incoherency in that pulse profile would result in a pulse duration getting longer than that in the ideal case (1), which is similar to the uncertainty relationship, $\delta E \times \delta \tau \geq \hbar / 2$, where the sign "=" corresponds to the ideal case of trans-spectral coherence (and even at that, only for a Gaussian profile); all the other cases correspond to the sign ">", and most often, to the sign "≫".  This can be readily illustrated by an incoherent radiation of a light bulb: having a huge spectrum, it can shine for hours.  By the same token, there is very little usefulness of observation of say $as$-pulses in a black-body radiation (as such an observation is claimed in some paper found in the literature on $as$-physics): the black-body radiation can be descried as the spectrum generated by huge numbers of supershort EM-pulses with random arrival times, durations, and shapes.  The pulses of sub-$fs$ and $as$ duration are plentiful in the sunlight, the only thing is they arrive and behave in a very random way.  The coherency and controllability are the major factors in the world of pulses.

While the sub-picosecond and femtosecond domains correspond to sub-eV energies, which are typical for molecular reactions, the domain below $0.15\ fs = 150\ as$ is the territory of atomic physics.  For example, the photoionization limit of the hydrogen atom, $13.6\ eV$,, is in the upper part of the spectrum of a $0.15\ fs$ pulse, which is about the time it takes for an electron at the ground state of the hydrogen atom to revolve around the proton.  A regular neutral atom, a maximum photoionization limit is about $24\ eV$ (in $He$).  Beyond the atomic-physics limits, the next in line are ions, especially those of heavy elements.  The larger the charge of a nucleus, and the fewer electrons left of the initially neutral atom, the more difficult it is to further ionize the ion. Going to the "ionic extremes", we can think of the heaviest stable atom, uranium, with all but one electron stripped away, by a high-intensity laser pulse, for example [17].  To remove that last electron, one needs more than $110 KeV$, close to the K-shell transition of uranium. This would take us into respectively shorter time scales of sub-attoseconds, $10^{-20}\ s = 0.01\ as$.  A potential application of those pulses and energies could be found e. g. in undestructive inspection of cargo containers and in medicine (one may notice that most used elements in medical hard X-ray tomography are *iodine* and *gadolinium* with their respective K-shell transition being being at $E_K = 33\,169\,KeV$ and $E_K = 50\,239\,KeV$ respectively. These are energies that correspond to time scale of ~ *10 zeptosecond* (see below).

Further beyond those energies is sort of a "quantum desert", in which no more atomic or ionic resonances can be found. Somewhere in the middle of it lies a border between regular (i.e. nonrelativistic) and relativistic quantum mechanics. It is determined by the rest-energy of an electron, $mc^2 \approx 0.5\,MeV$ (here *m* is the rest-mass of electron). Nonrelativistic quantum mechanics holds only for the energies significantly lower than $mc^2$, the spatial lengths and time intervals significantly longer than the so-called Compton wavelength, $\lambda_C = 2\pi\hbar/mc \approx 2.4 \times 10^{-2}$ Å, and Compton time $\hbar/mc^2 \approx 1.3 \times 10^{-21}\,s = 1.3\,zs$, respectively, where *zs* stands for a *zeptosecond* ($10^{-21}\,s$).

Thus, a sub-zeptosecond domain is a home for QED and nuclear processes. For example, the formation of (non-virtual) electron-positron pair and their annihilation are related to energies ~ $2\,mc^2 \approx 1\,MeV$ and times ~ $0.7\,zs$. The above-*1 Mev* domain also hosts other channels of electron-positron pair production, with the lowest threshold in such processes as $\gamma + N \rightarrow N + e^+ + e^-$, deuteron electro-disintegration, $e^- + {}^2H \rightarrow p + n + e^-$, and neutron photoproduction in beryllium $\gamma + {}^9Be \rightarrow n + {}^8Be$, and deiterium, $\gamma + {}^2H \rightarrow p + n$, with their thresholds from *1.2 MeV* to *2.2 MeV* (here $\gamma$ designates a high-energy photon).

Since in that domain, the atomic/ionic physics is left behind, one need to think about other ways of generating EM-spectra that reach the respective energies. The first proposal [18] to reach the *zs* domain was to use a "lasetron", based on the property of electrons driven by a circularly polarized powerful laser light to highly relativistic energies, to generate ultra-wide spectrum off radiation, exceeding the driving frequency by more than million times. The lasetron would generate *zs*-long EM bursts on a nuclear-time-scale using a Petawatt laser focused on solid particle or thin wire. The system would also generate pulse magnetic field up to ~ $10^6\,Tesla$. At the same time, the already widely available Terawatt lasers may generate *sub – attosecond* pulses of ~ $10^{-19}\,s$. The radiation by ultra-relativistic electrons driven by circularly-polarized high-intensity laser fields is basically reminiscent to synchrotron radiation; no synchrotron, however, can even come close to running electrons with the energy of *50 MeV* at the (laser) frequency $\omega_L \sim 10^{15} - 10^{16}\,s^{-1}$ in the *0.1 μm* radius orbit, as a Petawatt laser can. The lasetron can be achieved by placing a solid particle or a piece of wire of sub-wavelength cross-section in the focal plane of a super-powerful laser. A tight, sub-wavelength cloud of free electrons is formed then by the instant photoionization of target within the time much shorter the laser cycle; this cloud is driven by a circularly polarized laser in a $\lambda/\pi$-diameter circle with a speed close to the speed of light, and radiates a very narrow rotating cone of radiation [18] thus producing a hyper-short EM-burst at the point of observation. The Fourier spectrum of the bursts spreads up to the (classical) cutoff $\omega_{max} \sim 3\gamma^3 \omega_L$; here $\gamma = E_e/mc^2$ is a relativistic factor, with $E_e$ being a full energy of an electron. The major distinct feature here is the forced synchronization of the motion of all radiating electrons by the driving laser field. Radiation of such a synchronized bunch would be viewed by an observer in any point in the rotation plane as huge pulses/bursts of EM field as short as [19]

$$\tau_{pl} \sim 1/(2\omega_L \gamma^3) \, , \tag{2}$$

where $\gamma$ is the electron's relativistic factor. With $\lambda_L \equiv 2\pi c/\omega_L \sim 1\ \mu m$ and $\gamma \sim 64$ (attainable with a Petawatt laser), we have $\tau_{pl} \approx 1\ zs$; the high harmonics number here reaches $N_{HHG} = 3\gamma^3 \sim 0.8 \times 10^6$. For a laser with $10^{15}\ W$ (Petawatt) $\tau_{pl} = 2.6 \times 10^{-22}\ s\ (0.26\ zs)$. The classical cutoff of these bursts, $\hbar\omega_{cl} \approx 3\ MeV$, lies above the energy threshold of some photonuclear reactions discussed above. These numbers indicate the potential of lasetron bursts for time-resolved photonuclear physics.

In addition to zeptosecond pulses, the magnetic field at the center of rotation may reach $\sim 10^6\ Tesla$ -- comparable to fields in the vicinity of white dwarves. The driven motion of ionized electron cloud, essentially a strong current in a tight orbit, may create a strong magnetic (M) field normal to the rotation plane. The highest possible M-field in the lasetron can be estimated as [18]

$$B_{max} \sim e\,n_e\,\lambda_L/12 = \pi/6 \cdot (\lambda_L/\lambda_C) \cdot (n_e \cdot a_0^3) \cdot B_0 \tag{3}$$

where $n_e$ is the density of the cloud, $B_0 = e\alpha/a_0^2 \sim 1.33 \times 10^5\ G$ is the "Bohr" M-field scale, $a_0 \approx 0.53\ \text{Å}$, is Bohr radius and $\lambda_C$ is the Compton wavelength. Choosing a high-$Z$ electron-rich material we have $B_{max} \sim 4 \times 10^9\ G$ for $\lambda_L \sim 1\ \mu m$, and $\sim 4 \times 10^{10}\ G$ for $\lambda_L \sim 10\ \mu m$. The field will be oriented parallel to the laser propagation direction, and has the transverse size $\sim 2\rho \sim \lambda_L/\pi$; its duration will be about the same as that of the driving laser pulse.

A considerable amount of later theoretical and computational work was focused on the concept of zeptosecond pulses. It included the ionization of atoms by supershort pulses [20], pulses generation in terms of relativistic nonlinear Thomson scattering of an intense laser field by numerical simulations [21] and single *as* pulse generation [22], heat transport induced by *zs* pulses [23], modified Compton formula for frequency of photons by the scattering of $\delta$-like pulses [24], the spectrum of radiation backscattered from a laser by electrons [25], in the context of relativistic laser-plasma interactions [26], the radiation force on an electron [27], non-dipole transitions in atom excitation by ultrashort laser pulses [2Esc], application of Dirac equation for relativistic interaction of electron and supershort pulse [29], relation of $zs$ pulses other phenomena in optics in the relativistic regime [30], quasi-monochromatic x-rays from non-linear Thomson backscattering [31], and production of $zs$ pulses by the reflection of a relativistically intense $fs$ laser pulse from the boundary of an overdense plasma [32].

One of the related and greatly important area of research is to attain equally short electron beams ("e-bunches"). A recent research [33] on a fully relativistic theory of the field-gradient (ponderomotive) force (PF) in the ultra-high field in a standing laser wave showed that in addition to a large Kapitza-Dirac-effect, the PF in the direction normal to the incident momentum, exhibits a dramatic sign reversal with a relativistic threshold: While the remaining high-field repelling force along the incident momentum, in the direction normal to the incident momentum, PF reverses its sign and becomes a high-field attractive force if that momentum exceeds $mc$. The most interesting potentials exist in the case when laser has ultra-high field gradient and relativistic intensity. The presence of both these factors enables strong inelastic scattering of electrons crossing the laser beam [34]. This process allows for multi-MeV electron net acceleration per pass through a laser beam within the space less than a wavelength in a "laser-gate" configuration. It also allows for very tight temporal focusing *via* a klystron-like effect, and electron bunch formation down to a quantum, zepto-second limit even in low-gradient laser field. The ultimate limitation for time-focusing is imposed by uncertainty principle, due to which the finite e-bunch duration is:

$$\Delta\tau = (\hbar/mc^2)/\Delta\gamma \qquad (4)$$

where $\Delta\gamma$ is the maximum klystron-like modulation of the energy of the output E-beam due to laser driving; this limitation is similar to diffraction limit of focusing in optics. Using $\Delta\gamma \sim 0.2$ [34], and $\hbar\omega_L \sim 1\,eV$, we have $\Delta\tau \sim 30\,zs$, while an estimate based on the spread of E-beam with available lasers in a laser gate with energy spectrum $\Delta\tau \sim 45\,zs$, i. e. close to the quantum limit (4). Beyond the point of time focusing along the e-beam propagation axis, faster electrons over-run the slower ones, forming a shock wave, similar to that in a Coulomb explosions of clusters [35].

Finally, very recent work [36] proposed to attain yoctosecond ($1\,ys = 10^{-24}$) photon pulses from quark-gluon plasmas (QGP) based on the fact that experimentally, extremely short time scales can be reached through high energy collisions, especially during heavy-ion collisions that can produce QGP. By demonstrating that the emission envelope depends strongly on the internal dynamics of the QGP, it was shown in [36] that a double-peak structure potentially feasible in the emission envelope, could be a source for pump-probe experiments at the yoctosecond time scale. Such pulses could be used to time-resolved dynamics for example in baryon resonances, and provide an avenue to the study QGP dynamics during its expansion. It is worth noting that as was briefly mentioned in [35], the QGP expansion may produce shock wave, similar to those in Coulomb explosion.

Farther beyond heavy-ion collisions, there is the territory of high-energy physics with even larger energies; to give an idea of the time scales there, one may notice that a pulse with the highest photon energy of e. g. $1\,TeV$ (million of $MeV$s) could ideally be $\sim 10^{-27}\,s$ short. The shorter is a pulse (or the higher are the energies related to it), the farther away is the point in time we would be able to reach it. But once we are at that, a question is how far along this path can we travel at all [17]. The ultimate time scale of the quantum cosmology is $\sim 10^{-43}\,s$, the so called Planck time,

$$\tau_{Pl} = \sqrt{\hbar G/c^5} \approx 0.54 \times 10^{-43}\,s \tag{5}$$

(where $G \approx 6.7\,10^{-11}\,m^3/kg\,s^2$ is the gravitational constant), regarded as the time-scale of the birth-flash of the Big Bang, as well as an elementary "grain" or "pixel" of time, within which our "regular" physics of 4-D space+time breaks down into much greater number of dimensions hypothesized by the superstring theory (see e. g. [37]). A related Planck energy is $E_{Pl} \sim 1.2 \times 10^{28}\,eV = 1.2 \times 10^{16}\,TeV$. Beyond this time and energy scales, our understanding of time+space is getting a bit fuzzy. An intelligent being capable of controlling time and energy at those scales, can create a new universe (or destroy one), but following this line of thought, one can easily be accused of falling into a dangerous heresy (:-) of "intelligent design" concept, so let us say that the subject is outside the scope of this paper.

In conclusion, we browsed through a few ideas which can offer a potential for generating pulses a few orders of magnitude shorter than currently attained sub-femtosecond and attosecond pulses. They range from using highly-ionized atoms (sub-attoseconds), to electron bunches driven by relativistically-intense laser radiation (zeptoseconds), to quark-gluon plasmas in heavy-ion collisions (yoctoseconds). The field is wide open for new ideas both theoretical and experimental, and awaits for new explorations and explorers.

This work is supported by the USA Air Force Office of Scientific Research.

## References

[1]    R. L. Fork, C. H. Brito Cruz, P. C. Becker, and C. V. Shank, Opt. Lett., *12*, 483 (1987).

[2]    R. L. Carman, C. K. Rhodes, and R. F. Benjamin , Phys. Rev. A 24, 2649-2663 (1981)

[3]    S. M. Gladkov and N. I. Koroteev, Usp. Fiz. Nauk *160*, 105 (1990) [Sov. Phys. Usp. *33*, 554 (1990)];

[4]    G. Farkas and C. Toth, Phys. Lett. A *168*, 447 (1992);

[5]    T. W. Hänsch, Opt. Comm., *100*, 487 (1993).

[6]    P. B. Corkum, N. H. Burnett, and M. Y. Ivanov, Opt. Lett. *19*, 1870 (1994).

[7]    P. M. Paul, E. S. Toma, P. Breger, G. Mullot, F. Auge, Ph. Balcou, H. G. Muller, and P. Agostini, Science *292*, 1689, (2001).

[8]    M. Hentschel, R. Kienberger, Ch. Spielmann, G. A. Reider, N. Milosevic, T. Brabec, P. Corkum, U. Heinzmann, M. Drescher & F. Krausz, Nature *414*, 509 (2001)

[9]    M. Schultze, A. Wirth, I. Grguras, M. Uiberacker, T. Uphues, A. J. Verhoef, J. Gagnon, M. Hofstetter, U. Kleineberg, E. Goulielmakis, F. Krausz, Journal of electron spectroscopy and related phenomena *184*, 68 (2011).

[10]   A. E. Kaplan, Phys. Rev. Lett. *73*: 1243 (1994); A. E. Kaplan and P. L. Shkolnikov, JOSA B, *13*, 347 (1996)

[11]   R. W. Minck, R. W. Terhune, and W. G. Rado, Appl. Phys. Lett. 3, 181 (1963).

[12]   S. E. Harris and A. V. Sokolov, Phys. Rev. Lett, *81*, 2894 (1998); A. V. Sokolov, D. D. Yavuz, S. E. Harris, Opt. Lett., *24*,  557 (1999); F. Le Kien, J. Q. Liang, M. Katsuragawa, K. Ohtsuki, K. Hakuta, A. V. Sokolov, Phys.  Rev. A *60*,1562 (1999); A. Nazarkin, G. Korn, M. Wittmann, T. Elsaesser, Phys. Rev. Lett., *83*, 2560 (1999); A. V. Sokolov, D. R. Walker, D. D. Yavuz, G. Y. Yin, S. E. Harris, Phys. Rev. Lett., *85*, 562 (2000).

[13]   M. Y. Shverdin, D. R. Walker, D. D. Yavuz, G. Y. Yin, and S. E. Harris, Phys. Rev. Lett. *94*, 033904 (2005)

[14]   A. V. Sokolov, private communication.

[15]   A. H. Zewail, J. Phys. Chem. A, *104*,  5660 (2000)

[16]   A. E. Kaplan, J. Opt. Soc. Am. B (JOSA B), *15*: 951 (1998).

[17]   A. E. Kaplan, Nature, *431*, 633 (2004); also in "Optics & Photonics News (OPN), *17*, issue 2, 28 (2006)

[18] A. E. Kaplan and P. L. Shkolnikov, Phys. Rev. Lett. *88*, 074801 (2002); and follow-up discussion: W. R. Garrett, ibid. *89*, 279501 (2002); A. E. Kaplan and P. L. Shkolnikov, ibid. *89*, 279502 (2002); G. Stupakov and M. Zolotorev, ibid. *89*, 199501 (2002); A. E. Kaplan and P. L. Shkolnikov, ibid. *89*, 199502 (2002).

[19] L. Landau and E. Lifshitz, ***Classical Field Theory*** (Pergamon, New York, 1975), Ch. 48.

[20] V. I. Matveev, Techn. Phys. Lett., *28*, 874 (2002)

[21] K. Lee, Y. H. Cha, M. S. Shin, B. H. Kim, D. Kim, Phys. Rev. E *67*, 026502 (2003)

[22] P. F. Lan, P. X. Lu, W. Cao, Phys. of Plasmas, *13*, 013106 (2006)

[23] J. Marciak-Kozlowska, M. Kozlowski, Lasers in Engineering, *12*, 201 (2002)

[24] M. Pardy, Int. J. Theor. Phys. *42*, 99 (2003).

[25] F. He, Y. Y. Lau, D. P. Umstadter, R. Kowalczyk, Phys. Rev. Lett., *90*, 055002 (2003)

[26] D. P. Umstadter, J. Phys. D, *36*, R151-R165 (2003)

[27] J. A. Heras, Phys. Lett., A *314*, 272 (2003)

[28] A. V. Lugovskoy, I. Bray, J. Phys. B, *37* 3427 (2004);

[29] V. I. Matveev, E. S. Gusarevich, I. N. Pashev, J. Exper. Theor. Phys. *100*, 1043 (2005).

[30] G. A. Mourou, T. Tajima, S. V. Bulanov, Reviews of Modern Phys., *78*, 309 (2006)

[31] P. Lan, P. Lu, W. Cao, Phys. Scripta, *75*, 195 (2007)

[32] S. Gordienko, A. Pukhov, O. Shorokhov, and T. Baeva, Phys. Rev. Lett. *93*, 115002 (2004).

[33] A. E. Kaplan, A. L. Pokrovsky, Phys. Rev. Lett., *95*, 053601 (2005); A. L. Pokrovsky and A. E. Kaplan, Phys. Rev. A, *72*, 043401 (2005).

[34] A. E.Kaplan, A. L. Pokrovsky, Opt. Express, *17*, 6194 (2009).

[35] A. E. Kaplan, B. Y. Dubetsky, P. L. Shkolnikov, Phys. Rev. Lett., *91*, 143401, (2003)

[36] A. Ipp, H. C. Keitel, J. Evers, Phys. Rev. Lett., *103*, 152301 (2003)

[37] L. Smolin, ***Three Roads to Quantum Gravity***, Basic Books (2001); B. Greene, ***The Elegant Universe***, Random House Inc (2000).